\documentclass[aps,prd,amsmath,amssymb,showpacs,nofootinbib]{revtex4}
\usepackage{graphicx}

\usepackage{bm}

\newcommand{\beq}{\begin{equation}}
\newcommand{\eeq}{\end{equation}}
\newcommand{\bea}{\begin{eqnarray}}
\newcommand{\eea}{\end{eqnarray}}
\newcommand{\bseq}{\begin{subequations}}
\newcommand{\eseq}{\end{subequations}}

\newcommand{\Ref}[1]{(\ref{#1})}

\begin{document}

\title{Realistic cosmological scenario with non-minimal kinetic coupling}

\author{Sergey V. Sushkov$^{1,2}$} 
\email{sergey_sushkov@mail.ru}
\affiliation{$^1$ Institute of Mathematics and Mechanics and Institute of
Physics,
Kazan Federal University, Kremlevskaya str. 18, Kazan 420008, Russia\\
$^2$ Physics Department, CSU Fresno, Fresno, CA 93740-8031}

\begin{abstract}
We investigate cosmological scenarios in the theory of gravity with the scalar field possessing a non-minimal kinetic coupling to the curvature. It is shown that the kinetic coupling provides an essentially new inflationary mechanism.
Namely, at early cosmological times the domination of coupling terms in the field equations guarantees the quasi-De Sitter behavior of the scale factor: $a(t)\propto e^{H_{\kappa} t}$ with $H_\kappa=1/\sqrt{9\kappa}$, where $\kappa\simeq 10^{-74}$ sec$^2$ is the coupling parameter. The primary inflationary epoch driven by non-minimal kinetic coupling comes to the end at $t_f \simeq 10^{-35}$ sec. Later on, the matter terms are dominating, and the universe enters into the matter-dominated epoch which lasts approximately $0.5H_0^{-1}\sim 0.5\times10^{18}$ sec. Then, the cosmological term comes into play, and the universe enters into the secondary inflationary epoch with $a(t)\propto e^{H_{\Lambda} t}$, where $H_\Lambda=\sqrt{\Lambda/3}$.
Note that the present value of the acceleration parameter $q=\ddot a a/\dot a^2$ is estimated as $q_0\simeq0.25$, that is the universe is at the beginning of the epoch of accelerated expansion. Thus, the cosmological model non-minimal kinetic coupling represents the realistic cosmological scenario which successfully describes basic cosmological epochs and provide the natural mechanism of epoch change without any fine-tuned potential.

\end{abstract}

 \maketitle

\section{Introduction}

In recent decades the observational cosmology has been going through the period of the rapid growth. Precise measurements of the Cosmic Microwave Background (CMB) radiation \cite{CMB}, systematic observations of nearby and distant Type Ia supernovae (SNe Ia) \cite{supernova}, study of  baryon acoustic oscillations \cite{BAO}, mapping the large-scale structure of the Universe, microlensing observations, and many other remarkable results (see, for example, the review \cite{Observations}) have essentially expanded our knowledge about the Universe. Amazing discoveries, such as the accelerating expansion of the Universe and the dark matter evidence, have set new serious challenges before theoretical cosmology faced the necessity of radical modification of the standard model having successfully been exploited for a long time. Now, any viable cosmological model has to be able to describe several qualitatively different epoches of the Universe evolution, including the primary inflation, the matter-dominated stage, and the present acceleration (or secondary inflation). Moreover, it should also explain a mechanism providing an epoch change. These challenges have prompted many speculations mostly based on phenomenological ideas which involve new dynamical sources of gravity that act as dark energy, and/or various modifications to general relativity. The spectrum of models, having been postulated and explored in recent years, is extremely wide and includes, in particular, Quintessence \cite{quintessence}, K-essence \cite{Kessense}, Ghost Condensates \cite{Ghost}, Dvali-Gabadadze-Porrati gravity \cite{DGP}, $f(R)$ gravity \cite{fRgravity}, and others (see, e.g. Refs \cite{SahSta, PeeRat, Nob, CopSamTsu, CalKam, SilTro, Cli_etal, AmeTsu} for detailed reviews of these and other models).\footnote{This plethora of models reflects a deep crisis of phenomenological approach in the modern theoretical cosmology. To date, there are no unique criteria  to prefer one or another phenomenological model.}

It is worth noticing that the most of phenomenological models represents various modifications of scalar-tensor theories. For example, the quintessence is the ordinary scalar field with a fine-tuned potential; the K-essence is a scalar-tensor theory with a generalized kinetic term; the $f(R)$ gravity could be re-written in the Einstein frame as the ordinary general relativity with some effective scalar field, and so on. A broad class of cosmological models includes scalars non-minimally coupled to gravity \cite{Uzan:1999ch,nonminimal,liddle}. Additionally, one can further extend  scalar-tensor theories, allowing for non-minimal couplings between the derivatives of the scalar fields and the curvature \cite{Amendola}, and these scenarios reveal interesting cosmological and astrophysical behaviors \cite{Capozziello,kincoupl,Sus:2009,SarSus:2010, SusRom:2012}.




In our recent works \cite{Sus:2009,SarSus:2010} we investigated cosmological scenarios with the non-minimal kinetic coupling between the scalar field and the curvature, examining both the quintessence and the phantom cases in zero and constant potentials. According to the parameter choices and without the need for matter, we have obtained the variety of behaviors including a Big Bang, an expanding universe with no beginning, a cosmological turnaround, an eternally contracting universe, a Big Crunch, a Big Rip avoidance and a cosmological bounce. However, the most interesting and important feature we have found is that the non-minimal kinetic coupling provides an essentially new inflationary mechanism and naturally describe transitions between various cosmological phases without any fine-tuned potential. The inflation is driving by terms in the field equations responsible for the non-minimal kinetic coupling. At early times these terms are dominating, and the cosmological evolution has the quasi-de Sitter character $a(t)\sim e^{H_\kappa t}$, where $H_\kappa=1/\sqrt{9\kappa}$ and $\kappa$ is a coupling parameter. Later on, in the course of the cosmological evolution the domination of $\kappa$-terms is canceled, and this leads to a change of  cosmological epochs. In particular, in the cosmological scenario with zero potential the primary inflationary stage with $a(t)\sim e^{H_\kappa t}$ driving by non-minimal kinetic coupling changes into a power-law expansion with $a(t) \sim t^{1/3}$ \cite{Sus:2009}, whereas in the case of constant potential $V(\phi)\equiv\Lambda/8\pi$ the universe generally transits from one quasi-de Sitter phase to another with $a(t)\sim e^{H_\Lambda t}$, where $H_\Lambda=\sqrt{\Lambda/3}$ \cite{SarSus:2010}.

In this paper we consider more realistic cosmological models including an ordinary matter and the cosmological constant in addition to the scalar field with non-minimal kinetic coupling.

\section{Action and field equations\label{II}}

In this section we present the cosmological paradigm with non-minimal kinetic coupling between a scalar field and the curvature. In order to describe the quintessence and the phantom field in a unified way we adopt the $\varepsilon$-notation, that is the parameter $\varepsilon$ takes the value $+1$ for the canonical field and $-1$ for the phantom one.

\subsection{General fields equations with non-minimal derivative coupling}
Let us construct a gravitational theory of a scalar field $\phi$ with non-minimal derivative couplings to the curvature. In general one could have various forms of such couplings. For instance, in the case of four derivatives one could have the terms $\kappa_1 R\phi_{,\mu}\phi^{,\mu}$, $\kappa_2 R_{\mu\nu}\phi^{,\mu}\phi^{,\nu}$, $\kappa_3 R \phi\square\phi$, $\kappa_4 R_{\mu\nu} \phi\phi^{;\mu\nu}$, $\kappa_5 R_{;\mu} \phi\phi^{,\mu}$ and $\kappa_6 \square R \phi^2$, where the coefficients $\kappa_1,\dots,\kappa_6$ are coupling parameters with dimensions of length-squared. However, as it was discussed in \cite{Sus:2009,Amendola,Capozziello}, using total divergencies and without loss of generality, one can keep only the first two terms, $\kappa_1 R\phi_{,\mu}\phi^{,\mu}$ and $\kappa_2 R_{\mu\nu}\phi^{,\mu}\phi^{,\nu}$. Generally, field equations with non-minimal derivative coupling are of third order. However, in the specific case $-2\kappa_1=\kappa_2\equiv\kappa$ the coupling term $G_{\mu\nu} \phi^{,\mu}\phi^{,\nu}$ gives field equations of second order \cite{Sus:2009}.

Further we will consider cosmological scenarios in the theory of gravity with the action
\begin{equation}\label{action}
S=\int d^4x\sqrt{-g}\left\{ \frac{R}{8\pi} -\big[\varepsilon
g_{\mu\nu} + \kappa G_{\mu\nu} \big] \phi^{,\mu}\phi^{,\nu} -2
V(\phi)\right\}+S_m,
\end{equation}
where ${S}_m$ is the action for ordinary matter (not including the scalar field),
$V(\phi)$ is a scalar field potential, $g_{\mu\nu}$ is a metric, $R$ is the
scalar curvature, $G_{\mu\nu}$ is the Einstein tensor, and $\kappa$ is the
coupling parameter with dimension of ({\em length})$^2$. Varying the
action with respect to $g_{\mu\nu}$ and $\phi$ gives the field equations, respectively:
\bseq\label{fieldeq}
\bea
\label{eineq}
&& G_{\mu\nu}=8\pi\big[T_{\mu\nu}^{(m)}+T_{\mu\nu}^{(\phi)}
+\kappa \Theta_{\mu\nu}\big], \\
\label{eqmo}
&&[\varepsilon g^{\mu\nu}+\kappa G^{\mu\nu}]\nabla_{\mu}\nabla_\nu\phi=-V_\phi,
\eea
\eseq
where $V_\phi\equiv dV(\phi)/d\phi$, $T^{(m)}_{\mu\nu}$ is a stress-energy
tensor of ordinary matter, and
\bea \label{T}
T^{(\phi)}_{\mu\nu}&=&\varepsilon[\nabla_\mu\phi\nabla_\nu\phi-
{\textstyle\frac12}g_{\mu\nu}(\nabla\phi)^2]-g_{\mu\nu}V(\phi), \\
\Theta_{\mu\nu}&=&-{\textstyle\frac12}\nabla_\mu\phi\,\nabla_\nu\phi\,R
+2\nabla_\alpha\phi\,\nabla_{(\mu}\phi R^\alpha_{\nu)}
\nonumber\\
&&
+\nabla^\alpha\phi\,\nabla^\beta\phi\,R_{\mu\alpha\nu\beta}
+\nabla_\mu\nabla^\alpha\phi\,\nabla_\nu\nabla_\alpha\phi
\nonumber\\
&&
-\nabla_\mu\nabla_\nu\phi\,\square\phi-{\textstyle\frac12}(\nabla\phi)^2
G_{\mu\nu}
\label{Theta}\\
&&
+g_{\mu\nu}\big[-{\textstyle\frac12}\nabla^\alpha\nabla^\beta\phi\,
\nabla_\alpha\nabla_\beta\phi
+{\textstyle\frac12}(\square\phi)^2
\nonumber\\
&& \ \ \ \ \ \ \ \ \ \ \ \ \ \ \ \ \ \  \ \ \   \ \ \  \ \ \ \ \ \
\ \ \ \ \ -\nabla_\alpha\phi\,\nabla_\beta\phi\,R^{\alpha\beta}
\big]. \nonumber
\eea
Due to Bianchi identity $\nabla^\mu G_{\mu\nu}=0$ and the conservation law
$\nabla^\mu T^{(m)}_{\mu\nu}=0$, Eq. \Ref{eineq} leads to the differential
consequence
\beq
\label{BianchiT}
\nabla^\mu\big[T^{(\phi)}_{\mu\nu}+\kappa \Theta_{\mu\nu}\big]=0.
\eeq
Substituting Eqs. \Ref{T} and \Ref{Theta} into \Ref{BianchiT}, one can check straightforwardly that the differential consequence \Ref{BianchiT} is equivalent to \Ref{eqmo}.  In other words, Eq. \Ref{eqmo} is a differential consequence of Eq. \Ref{eineq}.

\subsection{Cosmological equations with non-minimal kinetic coupling}
Let us consider a spatially-flat FRW cosmological model with the metric
\begin{equation}
\label{metric} ds^2=-dt^2+a^2(t)d{\rm\bf x}^2,
\end{equation}
where $d{\rm\bf x}^2$ is the Euclidian metric, $a(t)$ is the scale factor, and $H(t)=\dot{a}(t)/a(t)$ is the Hubble parameter. Denoting the present moment of time as $t_0$, we have $a_0=a(t_0)$ and $H_0=H(t_0)$. Note that, without loss of generality, we can set $a_0=1$; in this case $a(t)$ is dimensionless. Supposing homogeneity and isotropy, we also get $\phi=\phi(t)$ and $T^{(m)}_{\mu\nu}=\mathop{\rm diag}(\rho,p,p,p)$, where $\rho=\rho(t)$ is the energy density and $p=p(t)$ is the pressure of matter.

Assume that the universe is filled with a perfect fluid consisting of several non-interacting components with the energy density $\rho_{i}$ and the pressure $p_{i}$. Each component obeys the conservation law $\nabla^\mu T^{(m_i)}_{\mu\nu}=0$, which in the metric \Ref{metric} takes the form of the continuity equation
\beq\label{continuity}
\dot\rho_{i}+3H(\rho_{i}+p_{i})=0.
\eeq
For a given barotropic equation of state $p_{i}=w_i\rho_{i}$ the continuity equation is easily integrated:
\beq
\rho_{i}=\rho_{{i}0} a^{-3(1+w_i)},
\eeq
where $\rho_{{i}0}$ is the energy density of the $i$-th component at $t=t_0$. The total energy density and pressure are found as
\bea\label{intrho}
\rho&=&\sum_i \rho_{i}=\sum_i \rho_{{i}0} a^{-3(1+w_i)}, \\
p   &=&\sum_i p_{i}=\sum_i w_i \rho_{{i}0} a^{-3(1+w_i)}.
\eea

The general field equations \Ref{fieldeq} written for the metric \Ref{metric}
yield
\bseq\label{genfieldeq}
\bea
  \label{00cmpt}
  &&3H^2=4\pi\dot{\phi}^2\left(\varepsilon-9\kappa H^2\right) +8\pi
V(\phi)+8\pi\rho,
  \\
  \label{11cmpt}
  &&\displaystyle
  2\dot{H}+3H^2=-4\pi\dot{\phi}^2
  \left[\varepsilon+\kappa\left(2\dot{H}+3H^2
+4H\ddot{\phi}\dot{\phi}^{-1}\right)\right]+8\pi V(\phi)-8\pi p,
  \\
  \label{eqmocosm}
  &&\varepsilon(\ddot\phi+3H\dot\phi)-3\kappa(H^2\ddot\phi
  +2H\dot{H}\dot\phi+3H^3\dot\phi)=-V_\phi,
\eea
\eseq
where a dot denotes derivatives with respect to time. Here it is worth noticing
that only two of the last three equations are independent, because Eq.
\Ref{eqmocosm} is a differential consequence of \Ref{00cmpt} and \Ref{11cmpt}.
Note also that Eq. \Ref{eqmocosm} can be represented in the more convenient form:
\beq\label{eqmoint}
\big[(\varepsilon-3\kappa H^2)\dot\phi\big]\!\dot{\phantom{\phi}}+
3H(\varepsilon-3\kappa H^2)\dot\phi=-V_\phi.
\eeq

\subsection{The case $V(\phi)\equiv{\rm const}$}
Hereafter we will discuss the particular case $V(\phi)\equiv{\rm const}$. In
fact, in this case the scalar potential plays the role of a cosmological
constant, and so we will use the notation
\beq
V(\phi)\equiv\frac{\Lambda}{8\pi},
\eeq
supposing that the cosmological constant $\Lambda$ is non-negative, i.e.
$\Lambda\ge0$. In the case $V(\phi)\equiv const$, when $V_\phi=0$, Eq. \Ref{eqmoint} can be easily integrated:
\beq\label{intphi}
\dot\phi=\frac{\varepsilon(2\lambda)^{1/2} }{a^3(\varepsilon-3\kappa H^2)},
\eeq
where $\lambda$ is a constant of integration.

Let us define the following dimensionless density parameters:
\beq
\Omega_{m_i0}=\frac{\rho_{i0}}{\rho_{cr}}, \quad
\Omega_{\phi0}=\frac{\lambda}{\rho_{cr}}, \quad
\Omega_{\Lambda0}=\frac{\Lambda}{8\pi\rho_{cr}},
\eeq
where $\rho_{cr}=3H_0^2/8\pi$ is the critical density. Using these notations and
taking into account Eqs. \Ref{intrho} and \Ref{intphi}, we can rewrite the
first-order equation \Ref{00cmpt} in terms of density parameters:
\beq\label{geneqH}
H^2=H_0^2\left[\Omega_{\Lambda0}+\sum_i\frac{\Omega_{m_i0}}{a^{3(1+w_i)}}+
\frac{\Omega_{\phi0}(\varepsilon-9\kappa H^2)}{a^6(\varepsilon-3\kappa H^2)^2}\right].
\eeq
At $t=t_0$ this equation reduces to
\beq
\Omega_{\Lambda0}+\sum_i\Omega_{m_i0}+ \frac{\Omega_{\phi0}(\varepsilon-9\kappa
H_0^2)}{(\varepsilon-3\kappa H_0^2)^2}=1.
\eeq
The latter represents a constraint for parameters $\Omega_{m_i0}$,
$\Omega_{\phi0}$, $\Omega_{\Lambda0}$.

\section{Cosmological scenarios with non-minimal kinetic coupling}
For given parameters, Eq.~\Ref{geneqH} completely determines the scale factor $a(t)$ and hence the whole cosmological evolution of the Universe. Hereafter, we will assume that $\varepsilon=1$, supposing the normal scalar field with positive kinetic energy. Also we will assume that the perfect fluid filling the Universe has only one non-relativistic component with zero pressure $p=0$ and the energy density $\rho=\rho_{0}a^{-3}$.


\subsection{Model with $\kappa=0$ and $\Lambda=0$.}
First we consider the simple model with $\kappa=0$ (no non-minimal kinetic coupling) and $\Lambda=0$ (no cosmological constant). In this case Eq. \Ref{geneqH} reads
\beq\label{k0L0}
H^2=H_0^2\left[\frac{\Omega_{m0}}{a^{3}}+\frac{\Omega_{\phi0}}{a^{6}}\right],
\eeq
with the constraint
\beq\label{cons1}
\Omega_{m0}+\Omega_{\phi0}=1.
\eeq
It is easy to see that $H\propto a^{-3}\to\infty$ if $a\to0$, and $H\propto a^{-3/2}\to0$ if $a\to\infty$.
Integrating Eq. \Ref{k0L0} together with the constraint \Ref{cons1} gives
\beq\label{ak0L0}
a(t)=\left[\textstyle\frac94\Omega_{m0}
H_0^2(t-t_0)^2+3H_0(t-t_0)+1\right]^{1/3},
\eeq
where $\Omega_{m0}\in[0,1]$. Note that the scale factor $a(t)$ becomes zero at the cosmological singularity $t_*=t_0-2(1-\sqrt{1-\Omega_{m0}})/3\Omega_{m0} H_0 <t_0$. Near the singularity
$a(t)\propto(t-t_*)^{1/3}$.
In the particular case $\Omega_{m0}=1$ one gets the well-known solution $a(t)=[\frac32 H_0(t-t_0)+1]^{2/3}$, describing the universe filled only with the dust. The opposite case $\Omega_{m0}=0$ corresponds to the universe filled only with the scalar field; in this case one finds $a(t)=[3H_0(t-t_0)+1]^{1/3}$. Note that the scalar field behaves effectively like the stiff matter with $p=\rho$. It is worth also noticing that, in fact, the scalar field could play the role of dark matter. Actually, first, it represents an additional gravitational source being equivalent to the perfect fluid (stiff matter). In particular, on cosmic scales the scalar field is slowing down the rate of the cosmological expansion. Second, it interacts only gravitationally with another matter, and so one can speculate that dark matter halos of galaxies are formed from the gravitating  scalar field.

\begin{figure}
\begin{center}
\parbox{5cm}{\includegraphics[width=5cm]{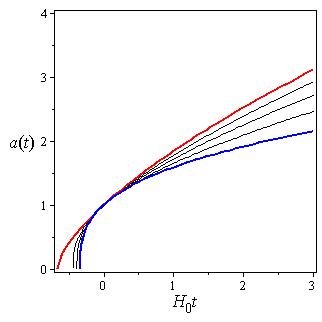}\\({\em a})}
\parbox{5.5cm}{\includegraphics[width=5.5cm]{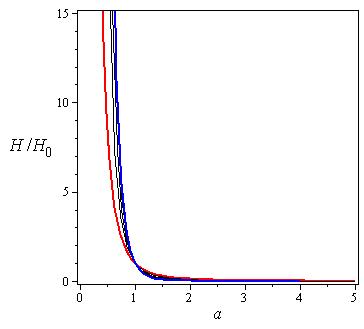}\\({\em b})}
\parbox{5cm}{\includegraphics[width=5cm]{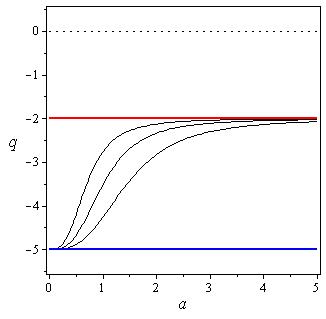}\\({\em c})}
\end{center}
\caption{Plots ({\em a}), ({\em b}), and ({\em c}) represent graphs for $a(t)$, $H(a)$, and $q(a)$, respectively, in the case $\kappa=0$ and $\Lambda=0$. Curves, from red to blue, are given for $\Omega_{\phi0}=0$; $0.25$; $0.5$; $0.75$; $1$, and $\Omega_{m0}=1-\Omega_{\phi0}$.
\label{fig1}}
\end{figure}

In Fig. \ref{fig1} we give plots of $a(t)$ for different values of $\Omega_{m0}$. It is seen that the less is $\Omega_{m0}$ and, in turn, the greater is $\Omega_{\phi0}=1-\Omega_{m0}$, the slower is the expansion rate. This feature explicitly demonstrates the role of the scalar field as a dark matter component.

Also, it is useful to represent graphically other cosmological characteristics: the Hubble parameter $H=\dot a/a$ and the acceleration parameter $q=\ddot a a/\dot a^2$, which is positive for accelerated and negative for decelerated expansion. The Hubble parameter $H$ is determined by Eq. \Ref{k0L0} as a function of $a$, i.e. $H(a)$. It is easy to check that $q(a)$ can be found as follows:
\beq\label{q}
q(a)=\frac{aH'(a)}{H(a)}+1,
\eeq
where $H'=dH/da$. Plots of $H(a)$ and $q(a)$ are shown in Fig. \ref{fig1}. Note that $q$ is negative for all values of $a$, i.e. the universe expands with deceleration in the course of the whole cosmological evolution.



Summarizing, one can conclude that the simple model considered above describes satisfactorily the matter-dominated stage of the universe evolution. Moreover, the scalar field component could be regarded as a candidate for dark matter. However, this model is unable to describe the inflation and the late-time acceleration of the universe.

\subsection{Model with $\kappa=0$ and $\Lambda>0$.}
Now we take into account a positive cosmological constant $\Lambda>0$ in addition to the ordinary matter and the minimally coupled scalar field. In this case Eq. \Ref{geneqH} yields
\beq\label{k0L>0}
H^2=H_0^2\left[\Omega_{\Lambda0}+\frac{\Omega_{m0}}{a^{3}}+\frac{\Omega_{\phi0}}{a^{6}}\right],
\eeq
with the constraint
\beq
\Omega_{\Lambda0}+\Omega_{m0}+\Omega_{\phi0}=1.
\eeq
In the limit $a\to0$ Eq. \Ref{k0L>0} reduces to \Ref{k0L0}. Therefore, the model with $\Lambda>0$ is also singular at $t=t_*$, and near the singularity one has $a(t)\propto(t-t_*)^{1/3}$. In the opposite limit $a\to\infty$ terms with $\Omega_{m0}$ and $\Omega_{\phi0}$ are negligibly small in comparison with $\Omega_{\Lambda0}$, and so one gets $H^2\approx H_0^2\Omega_{\Lambda0}=\Lambda/3$. The corresponding asymptotical solution has the quasi-De Sitter form $a(t)\propto e^{H_{\Lambda} t}$, where $H_{\Lambda}=\sqrt{\Lambda/3}$.

\begin{figure}
\begin{center}
\parbox{5cm}{\includegraphics[width=5cm]{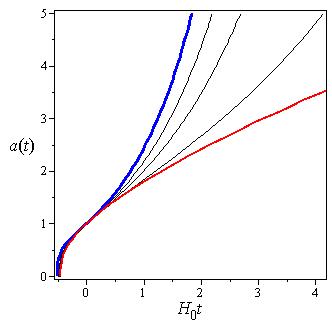}\\({\em a})}
\parbox{5.4cm}{\includegraphics[width=5.4cm]{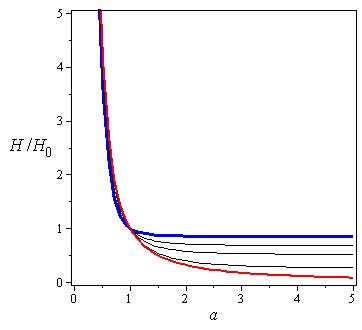}\\({\em b})}
\parbox{5.1cm}{\includegraphics[width=5.1cm]{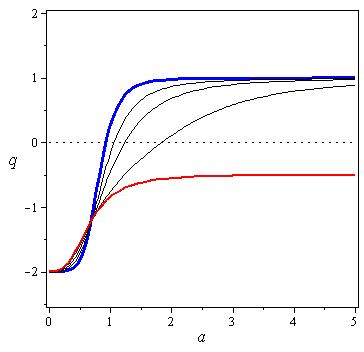}\\({\em c})}
\end{center}
\caption{Plots ({\em a}), ({\em b}), and ({\em c}) represent graphs for $a(t)$, $H(a)$, and $q(a)$, respectively, in the case $\kappa=0$ and $\Lambda>0$. Curves, from red to blue, are given for $\Omega_{\phi0}=0.23$, $\Omega_{\Lambda0}=0$; $0.07$; $0.27$; $0.47$; $0.73$, and $\Omega_{m0}=1-\Omega_{\phi0}-\Omega_{\Lambda0}$.
\label{fig2}}
\end{figure}

A general solution of Eq. \Ref{k0L>0} can be found in quadratures:
\beq
H_0(t-t_{0})=\int_{a_0=1} \frac{da}{a\left[\Omega_{\Lambda0}+\Omega_{m0}
a^{-3}+\Omega_{\phi0} a^{-6}\right]^{1/2}}.
\eeq
A graphical representation of this solution is shown in Fig.~\ref{fig2} for different sets of the density parameters $\Omega_{\Lambda0}$, $\Omega_{m0}$, and $\Omega_{\phi0}$. It is seen that at early stages the cosmological evolution has a matter-dominating character, whereas at late stages, when the cosmological constant is dominating, the universe expands with an acceleration. Fig.~\ref{fig2} represents also graphs for the Hubble parameter $H=\dot a/a$ and the acceleration parameter $q=\ddot a a/\dot a^2$ as functions of $a$. The behavior of $q$ demonstrates that the universe expands with deceleration during the matter-dominated epoch, when $q$ is negative; then $q$ becomes positive and the deceleration is changed into acceleration signifying the beginning of new cosmological epoch. At the moment of epoch change the acceleration parameter is equal to zero, i.e. $q=0$. Note that this moment depends on a ratio between the density parameters. In particular, the greater is $\Omega_{\Lambda0}$ the earlier is the beginning of the accelerating phase (see Fig. \ref{fig2}).

So, we can conclude that the model with a cosmological constant (or a constant scalar field potential) describes satisfactorily the matter-dominated stage and the late-time acceleration of the universe. However, the model is still unsuitable for describing an initial inflationary phase.

\subsection{Model with $\kappa>0$ and $\Lambda>0$}
Now, let us discuss the role of non-minimal kinetic coupling in the cosmological evolution. Assuming $\kappa>0$ in Eq. \Ref{geneqH}, we find
\beq\label{k_not=0L>0}
H^2=H_0^2\left[\Omega_{\Lambda0}+\frac{\Omega_{m0}}{a^{3}} +\frac{\Omega_{\phi0}(1-9\kappa H^2)}{a^{6}(1-3\kappa H^2)^2}\right],
\eeq
with the constraint
\beq\label{constraint3}
\Omega_{\Lambda0}+\Omega_{m0}+\Omega_{\phi0}\frac{1-9\kappa H_0^2}{(1-3\kappa
H_0^2)^2}=1.
\eeq
In the limit $a\to\infty$ terms with $\Omega_{m0}$ and $\Omega_{\phi0}$ become negligibly small, so that Eq. \Ref{k_not=0L>0} takes the asymptotical form $H^2\approx H_0^2\Omega_{\Lambda0}=\Lambda/3$, and the corresponding asymptotical solution has the quasi-De Sitter form $a(t)\propto e^{H_{\Lambda} t}$.
Note that it does not depend on the coupling parameter $\kappa$ at all. The essential role of non-minimal kinetic coupling manifests itself only at early stages. Actually, in the limit $a\to0$ the Hubble parameter has the following asymptotical behavior (see the appendix):
\beq
H=\sqrt{1/9\kappa} +O(a^3).
\eeq
Therefore, the early-time ($t\to-\infty$) cosmological evolution has the
quasi-De-Sitter (inflationary) behavior with $a(t)\propto e^{H_{\kappa} t}$, where $H_{\kappa}=\sqrt{1/9\kappa}$.

\begin{figure}
\begin{center}
\parbox{5cm}{\includegraphics[width=5cm]{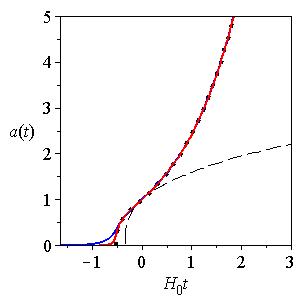}\\({\em a})}
\parbox{5.3cm}{\includegraphics[width=5.3cm]{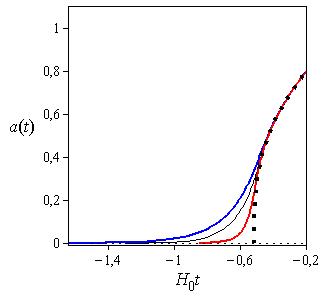}\\({\em b})}\\
\parbox{5.3cm}{\includegraphics[width=5.3cm]{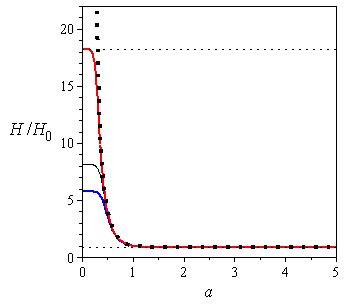}\\({\em c})}
\parbox{4.9cm}{\includegraphics[width=4.9cm]{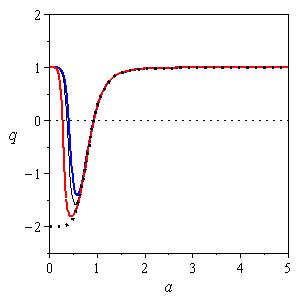}\\({\em d})}
\end{center}
\caption{Plots ({\em a}) and ({\em b}) represent graphs for $a(t)$ in different scales; plots ({\em c}) and ({\em d}) represent graphs for $H(a)$ and $q(a)$, respectively, in the case $\kappa>0$ and $\Lambda>0$. Curves, from red to blue, are given for $\gamma\equiv 3\kappa H_0^2=0.001; 0.005; 0.01$, $\Omega_{m0}=0.04$ and $\Omega_{\phi0}=0.23$; the value of $\Omega_{\Lambda0}$ obeys the constraint \Ref{constraint3}. The dashed curve in ({\em a}) corresponds to $\kappa=0$, $\Omega_{\Lambda0}=0$, $\Omega_{m0}=0$, $\Omega_{\phi0}=1$;  dotted curves correspond to $\kappa=0$, $\Omega_{\Lambda0}=0.73$, $\Omega_{m0}=0.04$,  $\Omega_{\phi0}=0.23$.
\label{fig3}}
\end{figure}

A general solution of Eq. \Ref{k_not=0L>0} can be found numerically. In Figs.~\ref{fig3}{\em a,b} we represent plots of $a(t)$ given for different values of $\kappa$. In addition, we give graphs for the Hubble parameter $H=\dot a/a$ and the acceleration parameter $q=\ddot a a/\dot a^2$ as functions of $a$ (see Fig.~\ref{fig3}{\em c,d}). Graphical solutions explicitly demonstrate that a cosmological scenario in the model with the non-minimal kinetic coupling, positive cosmological constant and matter has three qualitatively different stages.  At early times $-\infty<t<t_1$ the Universe is exponentially expanding, so that $a(t)\propto e^{H_{\kappa} t}$, $H\propto H_{\kappa}$, and $q\propto 1$.  It is necessary to stress that the initial inflationary stage is completely provided with the non-minimal kinetic coupling of the scalar field to the curvature; the matter and cosmological constant do not play any role in this epoch. Then, at $t_1<t<t_2$, the Universe expansion takes a power-law character. This is a matter-dominated epoch with the negative acceleration parameter $q$ (see Fig. \ref{fig3}). At late times $t>t_2$ the positive cosmological constant becomes dominating and the Universe expansion again takes the inflationary character such that $a(t)\propto e^{H_{\Lambda} t}$, $H\propto H_{\Lambda}$, and $q\propto 1$.

\subsection{Estimations}
To estimate parameters of the phenomenological model constructed above, we will use well-established cosmological facts. As is known, the initial inflationary epoch should be over
at $t_f \simeq 10^{-35}$ sec, and it should last {at least} 60 Hubble times (e-folds) (see, for example, Ref. \cite{Mukhanov}). Since $a(t)\propto e^{H_\kappa t}$ during this epoch, we have $H_\kappa t_f\sim 60$, or $H_\kappa\simeq 6\times10^{36}$ sec$^{-1}$. Then, taking into account that $H_\kappa=\sqrt{1/9\kappa}$, we find $\kappa\simeq 10^{-74}$ sec$^2$, or $l_\kappa\simeq 10^{-27}$ cm, where $l_\kappa=\kappa^{1/2}$ is the corresponding non-minimal coupling length.

The present value of the Hubble parameter is $H_0\sim 70\ {\rm (km/sec)/ Mpc}\sim 10^{-18}$ sec$^{-1}$, hence we obtain $\gamma=3\kappa H_0^2\simeq 10^{-109}$. Note that the value of $\gamma$ is extremely small in comparison with unity, and so neglecting terms with $\gamma$ in the constraint \Ref{constraint3} yields $\Omega_{\Lambda0}+\Omega_{m0}+\Omega_{\phi0}=1$. It is reasonable to assume $\Omega_{\Lambda0}=\Omega_{DE}$, $\Omega_{\phi0}=\Omega_{DM}$, $\Omega_{m0}=\Omega_{b}$, where  $\Omega_{DE}\simeq 0.73$ is the dark energy density, $\Omega_{DM}\simeq 0.23$ is the dark matter density, and $\Omega_{b}\simeq 0.04$ is the baryon density. With these values in hand we can calculate the acceleration parameter $q(a)$ at the present time $t_0$ with $a(t_0)=a_0=1$. Substituting $\Omega_{\Lambda0}=0.73$, $\Omega_{\phi0}=0.23$, $\Omega_{m0}=0.04$ into Eq. \Ref{q}, we find $q_0=q(a_0)=0.25$. It is worth noticing that $q_0$ is positive, that is the model with given density parameters describes the present cosmic acceleration. Stress also that $q_0<1$, what means the beginning of the secondary accelerated epoch. Later on, the accelerated expansion will enter into the quasi-De Sitter phase with $q\sim 1$.

With given $\Omega_{\Lambda0}=0.73$, $\Omega_{\phi0}=0.23$, and $\Omega_{m0}=0.04$ we can also estimate the duration of matter-dominated phase, i.e. $t_{MD}$. Supposing that $a=0$ at the beginning and $a=1$ at the end of the phase, we find
$$
H_0 t_{MD}=\int_0^1 \frac{da}{a\left[\Omega_{\Lambda0}+\Omega_{m0}
a^{-3}+\Omega_{\phi0} a^{-6}\right]^{1/2}}\simeq 0.5.
$$
Hence, $t_{MD}=0.5 H_0^{-1}\sim 0.5\times 10^{18}$ sec.

\section{Summary and conclusion}
In this paper we have explored a cosmological model with the scalar field possessing non-minimal kinetic coupling to the curvature given as $\kappa G_{\mu\nu}\phi^{,\mu}\phi^{,\nu}$, where $\kappa$ is a coupling parameter with dimension of ({\em length})$^2$. Additionally, the model includes an ordinary matter in the form of a pressureless perfect fluid, and the positive cosmological constant $\Lambda$, which could be also considered as the constant scalar potential $V(\phi)\equiv \Lambda/8\pi$. It is worth noticing that two dimensional parameters $\kappa$ and $\Lambda$ determine, eventually, two characteristic scales $H_\kappa=1/\sqrt{9\kappa}$ and $H_\Lambda=\sqrt{\Lambda/3}$.

The considered model represents the number of interesting and important features.
\begin{itemize}
\item
The most important feature is that the non-minimal kinetic coupling provides an {\em essentially new} inflationary mechanism which does not need any fine-tuned potential. The essence of the mechanism is that at early cosmological times the coupling $\kappa$-terms in the field equations are dominating and provide the quasi-De Sitter behavior of the scale factor: $a(t)\propto e^{H_{\kappa} t}$ with $H_\kappa=1/\sqrt{9\kappa}$. Assuming that the initial inflationary epoch should be over at $t_f \simeq 10^{-35}$ sec, and it should last 60 Hubble times (e-folds), one can estimate the coupling parameter $\kappa\simeq 10^{-74}$ sec$^2$ and the corresponding length $l_\kappa\equiv\kappa^{1/2}\simeq 10^{-27}$ cm.

\item
The cosmological scenario consists of three basic epochs: the primary inflation, the matter-dominated phase, and the secondary inflation. The primary inflationary epoch driven by non-minimal kinetic coupling comes to the end at $t_f \simeq 10^{-35}$ sec. Later on, the matter terms in the field equations begin to be dominating, and the universe enters into the matter-dominated epoch which is characterized by decelerated expansion. This epoch lasts approximately $0.5H_0^{-1}\sim 0.5\times10^{18}$ sec, then the cosmological term in the field equations comes into play, and the universe enters into the secondary inflationary epoch with $a(t)\propto e^{H_{\Lambda} t}$, where $H_\Lambda=\sqrt{\Lambda/3}$. Note that the acceleration parameter $q=\ddot a a/\dot a^2$ during the inflation is $q\sim 1$. The present value of $q$ is estimated as $q_0\simeq0.25$. This means that at present the universe is at the beginning of the epoch of accelerated expansion.

\item
The model provides a natural mechanism of epoch change.  Mathematically, every cosmological epoch corresponds with an appropriate term dominating in field equations in a definite time interval, and a change of epochs occurs when one dominating term is replaced by another. In this connection, it is necessary to stress that now one needs no reheating mechanism to create matter after the primary inflation. Actually, the ordinary matter exists during the whole period of the inflation driven by non-minimal kinetic coupling, however its influence on the cosmological evolution is suppressed.

\end{itemize}

Summarizing, we can conclude that  the theory of gravity with the scalar field possessing the non-minimal kinetic coupling to the curvature yields realistic cosmological scenarios which successfully describe basic cosmological epochs and provide the mechanism of epoch change.

\section*{Acknowledgments}
The work was supported in part by the Russian Foundation for Basic Research grants No. 11-02-01162. I appreciate California State University Fresno and personally Douglas Singleton for hospitality during the Fulbright scholarship visit. Also, I am grateful to Sergey Rubin for useful discussions.

\section*{Appendix}
Here we will discuss in details the equation \Ref{k_not=0L>0} which reads
\beq\label{app1}
H^2=H_0^2\left[\Omega_{\Lambda0}+\frac{\Omega_{m0}}{a^{3}}
+\frac{\Omega_{\phi0}(1-9\kappa H^2)}{a^{6}(1-3\kappa H^2)^2}\right],
\eeq
Using new notations
\bea
&& y=(H/H_0)^2,\quad x=a^3, \quad \gamma=3\kappa H_0^2,\nonumber\\
&&\omega_0=\Omega_{\Lambda0},\quad \omega_1=\Omega_{m0},\quad
\omega_2=\Omega_{\phi0},
\label{not}
\eea
we can rewrite it as follows:
\beq\label{app2}
y=\omega_{0}+\frac{\omega_{1}}{x} +\frac{\omega_{2}} {x^2}\frac{1-3\gamma
y}{(1-\gamma y)^2}.
\eeq
The latter is a cubic equation for  $y(x)$. Let us discuss asymptotical properties of $y(x)$. In the limit $x\to\infty$ we can neglect two last terms in Eq. \Ref{app2} and find
\beq\label{asinfty}
y=\omega_0+O(x^{-1}).
\eeq
To obtain asymptotics in the limit $x\to0$, we represent $y(x)$ as follows:
$$
y(x)=\sum_{k=1}^\infty \frac{y_{-k}}{x^k}+\sum_{n=0}^\infty y_n x^n.
$$
Substituting this series straightforwardly into Eq. \Ref{app2} yields three
different asymptotical solutions at $x\to0$:
\bseq\label{xto0}
\bea
\label{aszero}
y^{(1)}&=&\frac{1}{3\gamma}+\frac{4\omega_{1}}{27\gamma\omega_{2}}x-\frac{4}{243
}
\frac{3\omega_{2}+4\gamma\omega_{1}^2-9\gamma\omega_{0}\omega_{2}}{
\gamma^2\omega_{2}^2}x^2 +O(x^3),\\
\label{y23}
y^{(2,3)}&=&
\frac{\gamma\omega_{1}\pm\sqrt{\gamma^2\omega_{1}^2-12\gamma\omega_{2}}}{2\gamma
x}+O(x^0).
\eea
\eseq
Note that the first solution $y^{(1)}$ has a regular behavior near $x=0$;
namely, $y^{(1)}\to1/3\gamma$ if $x\to0$. It is worth noticing that the
asymptotical value $y^{(1)}(0)=1/3\gamma$ does not depend on the parameters
$\omega_i=\{\omega_0,\omega_1,\omega_2\}$ at all, and so the asymptotic
$y^{(1)}\approx 1/3\gamma$ at $x\to0$ has the universal character for Eq.
\Ref{app2}. Note also that the solution $y^{(1)}$ is real for any real values of
$\omega_{i}$. On the contrary, the solutions $y^{(2,3)}$ are singular at $x=0$;
their asymptotics are $y^{(2,3)}\propto 1/x$ at $x\to0$. Moreover, $y^{(2,3)}$
could be real or complex conjugate depending on a sign of the operand of the
square root in Eq. \Ref{y23}.

To obtain a general solution, we rewrite Eq. \Ref{app2} in an equivalent form:
\beq\label{cubeeq}
y^3+p_2 y^2+p_1 y + p_0=0,
\eeq
where
\bseq\label{p}
\bea
p_2&=&-\frac{(2+\gamma\omega_0)x+\gamma\omega_1}{\gamma x},\\
p_1&=&\frac{(1+2\gamma\omega_0)x^2+2\gamma\omega_1 x+3\gamma\omega_2}{\gamma^2
x^2},\\
p_0&=&-\frac{\omega_0 x^2 + \omega_1 x + \omega_2}{\gamma^2 x^2}
\eea
\eseq
are real. A cubic function $y^3+p_2 y^2+p_1 y + p_0$ with real coefficients $p_i$
has generally three roots given by Cardano's formulae (see, for example,
\cite{KornKorn}):
\bseq\label{solcubeeq}
\bea
\label{y1}
y^{(1)} &=& y^{(1)}(x;\gamma,\omega_i) \equiv
-\frac{p_2}{3}-\frac{C}{3}-\frac{p_2^2-3p_1}{3C},\\
y^{(2,3)} &=& y^{(2,3)}(x;\gamma,\omega_i) \equiv -\frac{p_2}{3}+\frac{C(1\pm
i\sqrt{3})}{6} +\frac{(1\mp i\sqrt{3})(p_2^2-3p_1)}{6C},
\eea
\eseq
where
\bseq\label{CQ}
\bea
C&=&\sqrt[3]{\textstyle\frac12(Q+2p_2^3-9p_2p_1+27p_0)},\\
Q&=&\sqrt{(2p_2^3-9p_2p_1+27p_0)^2-4(p_2^2-3p_1)^3}.
\eea
\eseq
Note that the above expressions are not yet unambiguously defined, and so they
need further explanation. To do this, we will assume that the square root refers
to the principal (positive) square root when its operand is non-negative, and
the cube root is to be interpreted as the real one, so that the solution
$y^{(1)}$ is real, and $y^{(2,3)}$ are complex conjugate. Otherwise, there is no
real square root and we can arbitrarily choose one of the imaginary square
roots. Then, for extracting the complex cube roots of the resulting complex
expression, we will choose among three cube roots that which provides $y^{(1)}$
to be real.

Since the root $y^{(1)}$ is real, we will consider it as a physical solution of the cubic
equation \Ref{cubeeq}. It is necessary to emphasize that $y^{(1)}$ has the
regular asymptotic \Ref{xto0} in the limit $x\to0$ (one can check this fact
straightforwardly). Now, reverting to the old notations, we find
\beq\label{H}
H^2=H_0^2 y^{(1)}(a;3\kappa H_0^2,\Omega_{\Lambda0},\Omega_{m0},\Omega_{\phi0}).
\eeq
To obtain an explicit expression
for $H$, one should use relations \Ref{p}, \Ref{solcubeeq}, \Ref{CQ}. However, a
final expression is too cumbersome, and so we skip this step. Instead, we will
discuss an asymptotical behavior of $H$. Taking into account Eqs. \Ref{asinfty},
\Ref{aszero}, we can easily find
\bea
&& H=\sqrt{\Lambda/3} +O(a^{-3}) \mbox{\rm ~~if~~} a\to\infty,\\
&& H=\sqrt{{1}/{9\kappa}} +O(a^3) \mbox{\rm ~~if~~} a\to0.
\eea
Finally, integrating Eq. \Ref{H} yields
\beq
H_0(t-t_0)=\int_{a_0=1}\frac{da/a}{[y^{(1)}(a;3\kappa H_0^2,\Omega_{\Lambda0},\Omega_{
m0},\Omega_{\phi0})]^{1/2}}.
\eeq
This relation could be used for numerical calculations of $a(t)$.


\end{document}